# Decomposition of Time Series Data of Stock Markets and its Implications for Prediction – An Application for the Indian Auto Sector

Jaydip Sen
Calcutta Business School, Diamond Harbor Rod, Bishnupur – 743503
West Bengal, INDIA
email: jaydips@calcuttabusinessschool.org

and

Tamal Datta Chaudhuri
Calcutta Business School, Diamond Harbour Road, Bishnupur – 743503
West Bengal, INDIA
email: tamalc@calcuttabusinessschool.org

**ABSTRACT**

With rapid development and evolution of sophisticated algorithms for statistical analysis of time series data, the research community has started spending considerable effort in technical analysis of such data. Forecasting is also an area which has witnessed a paradigm shift in its approach. In this work, we have used the time series of the index values of the Auto sector in India during January 2010 to December 2015 for a deeper understanding of the behavior of its three constituent components, e.g., the Trend, the Seasonal component, and the Random component. Based on this structural analysis, we have also designed three approaches for forecasting and also computed their accuracy in prediction using suitably chosen training and test data sets. The results clearly demonstrate the accuracy of our decomposition results and efficiency of our forecasting techniques, even in presence of a dominant Random component in the time series.

**Key words:** Decomposition, Trend, Seasonal, Random, Holt Winters Forecasting model, Neural Network, Back Propagation Network, ARIMA, VAR, Bayesian Vector Autoregressive (BVAR) model.

**JEL Classification:** G 11, G 14, G 17, C 63

## 1. INTRODUCTION

The Efficient Stock Market Hypothesis has generated significant interest in the literature and there has been research work trying to prove and disprove it. Whether stock prices can be predicted, has been keenly debated, and although the randomness of stock price movements has never been doubted, efforts have been made to throw light on ways to predict stock prices. In their seminal book, Graham and Dodd [1] tried to educate the investors about ways to pick stocks and avoid pitfalls. They didn't attempt to predict stock price movements, but enlightened us on parameters to look out for while picking stocks or sectors. Murphy [2] explicitly laid down the principles of technical analysis of stock prices and pointed out the differences between fundamental analysis and technical analysis. The objective was to show that stock picking, buying and selling, can be formalized in terms of various technical indicators and distinct patterns can be identified which can aid in stock price prediction. Various publications in India, like Economic Times Wealth Plus, Dalal Street and Business Standard, and also various Business TV channels continuously provide advice on how to pick stocks for a healthy portfolio. They tell us to look for sectoral characteristics, market timing, period of holding, and fundamentals of companies.

One of the important factors in selecting a stock of a company, apart from fundamentals and management quality, is the sector to which the company belongs. Various sectors display varying characteristics and it is important to understand the sectoral patterns.

## 2. OBJECTIVE OF THE STUDY

The purpose of this paper is to breakdown time series data of sectoral indices into trend, seasonal and random components. This will help in stock selection in the following ways. First, it will indicate the overall trend of the sector, hence the stock price, and help in taking a position. Second, if seasonality patterns can be seen, then during which month which sector and hence which stock should be a good buy, can be inferred. Third, the random component will throw some light on the volatility pattern of the sector and hence the stock. This decomposition will indicate which of the three components are stronger and can shed further light on the efficient market hypothesis.

This decomposition can also throw some light on certain beliefs, like i) the small cap sector is more random, hence speculative; ii) the healthcare sector is less random; iii) the Auto and FMCG sector is seasonal; iv) the capital goods sector is tied to the India growth story; and v) the IT sector is tied to the world growth story. The decomposition will bring out the overall macroeconomic characteristic of a sector, which affects the fundamentals of a company.

The rest of the paper is organized as follows. Section 3 briefly discusses the methodology in constructing various time series and decomposing the time series into its components. Section 4 presents the results of decomposition of the auto sector index time series values into Trend, Seasonal and Random components. Inferences are made on the roles played by three components on the overall time series index values. Section 5 presents four forecasting approaches and one approach for analyzing the behavior of the structural components of the auto sector index time series. Section 6 presents some related work in the current literature. Finally, Section 7 concludes the paper.

## 3. METHODOLOGY

In this work, we have used daily index data for the Auto sector for the period January 2010 to December 2015. The daily index values are first aggregated into monthly averages resulting into 70 values in each time series data. We use R libraries to convert each of these raw data series into a monthly time series. Each of these R time series now is an aggregation of three components: (i) Trend, (ii) Seasonal, and (iii) Random. In order to make further investigations into the behavior of the time series data, we decompose each time series into its three components. We use R library "TTR" for this purpose. After the decomposition, the behavior of the index, on the values of the components, are analysed. We also apply some robust forecasting techniques on this data and critically analyze the accuracy of the forecasting methods that we have applied.

## 4. DECOMPOSITION RESULTS

In this Section, we present some of the results that we have obtained in time series decomposition work. We particularly focus on the Auto sector time series values and discuss the results that we have obtained from its decomposition.

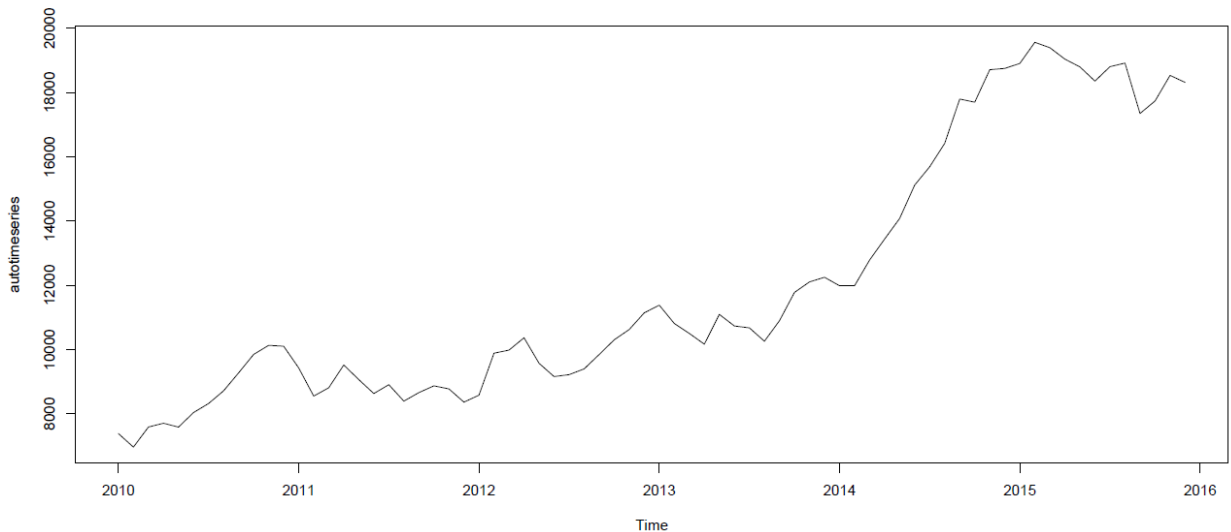

**Fig 1: Auto index time series (Jan 2010 – Dec 2015)**

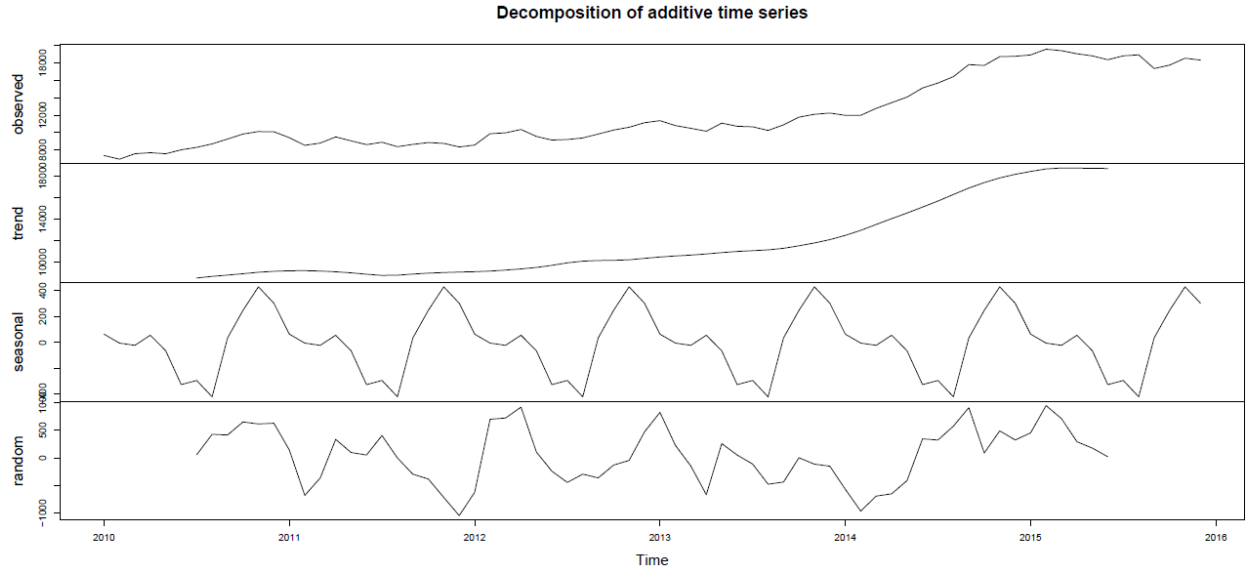

**Fig 2: Decomposition of auto index time series into trend, seasonal and random components**

Figure1 shows the overall time series for the Auto sector index for the period January 2010 – December 2015. It is not difficult to see that the time series has an increasing trend till the curve has exhibited as small downward fall during the latter part of 2015. Figure 2 shows the decomposition result of the time series in Figure 1. The three components of the time series are shown separately so that their relative behavior can be visualized.

Table 1 presents the numerical values of the time series data and all its three components. The Trend and the random components are not available for the period January 2010 – June 2010 and also during the period July 2015 – December 2015. This is due to the fact that Trend computation requires some long term data. In order to compute Trend figures for January 2010 – June 2010 we need time series data from July 2009 – December 2009 (which is not available in our raw dataset). In the same line, for computing Trend figures for July 2015- December 2015, time series data from January 2016 – June 2016 are needed. Due to the non-availability of the Trend values for these periods, it is not possible to compute the Random components too. Since the aggregate of the Trend, Seasonal and Random components is the final time series figure and because of the fact that Seasonal components remain constant for the same month over the period, the absence of Trend values makes it impossible for us to compute the Random components for these specific months.

**Table 1: Auto Sector Index Time Series and its Components**

| Year | Month | Time Series Aggregate | Trend | Seasonal | Random |
|---|---|---|---|---|---|
| 2010 | January | 7380 | | 63.611111 | |
| | February | 6958 | | -5.280556 | |
| | March | 7584 | | -22.672222 | |
| | April | 7702 | | 55.419444 | |
| | May | 7581 | | -65.030556 | |
| | June | 8034 | | -325.772222 | |
| | July | 8315 | 8552.250 | -293.705556 | 56.455556 |
| | August | 8710 | 8703.708 | -419.822222 | 426.113889 |
| | September | 9269 | 8820.833 | 34.677778 | 413.488889 |
| | October | 9844 | 8947.292 | 247.344444 | 649.363889 |
| | November | 10127 | 9084.500 | 428.969444 | 613.530556 |
| | December | 10100 | 9170.833 | 302.261111 | 626.905556 |
| 2011 | January | 9426 | 9219.958 | 63.611111 | 142.430556 |
| | February | 8547 | 9231.083 | -5.280556 | -678.802778 |
| | March | 8806 | 9192.208 | -22.672222 | -363.536111 |
| | April | 9515 | 9125.917 | 55.419444 | 333.663889 |
| | May | 9061 | 9028.667 | -65.030556 | 97.363889 |
| | June | 8626 | 8899.625 | -325.772222 | 52.147222 |
| | July | 8902 | 8791.667 | -293.705556 | 404.038889 |
| | August | 8390 | 8811.917 | -419.822222 | -2.094444 |
| | September | 8656 | 8916.458 | 34.677778 | -295.136111 |
| | October | 8866 | 9000.667 | 247.344444 | -382.011111 |
| | November | 8771 | 9057.125 | 428.969444 | -715.094444 |
| | December | 8359 | 9100.250 | 302.261111 | -1043.511111 |
| 2012 | January | 8576 | 9135.292 | 63.611111 | -622.902778 |
| | February | 9883 | 9190.167 | -5.280556 | 698.113889 |
| | March | 9979 | 9281.375 | -22.672222 | 720.297222 |
| | April | 10363 | 9390.458 | 55.419444 | 917.122222 |
| | May | 9568 | 9527.208 | -65.030556 | 105.822222 |
| | June | 9154 | 9720.083 | -325.772222 | -240.311111 |
| | July | 9215 | 9952.708 | -293.705556 | -444.002778 |
| | August | 9394 | 10108.083 | -419.822222 | -294.261111 |
| | September | 9841 | 10168.333 | 34.677778 | -362.011111 |
| | October | 10299 | 10181.708 | 247.344444 | -130.052778 |
| | November | 10620 | 10236.875 | 428.969444 | -45.844444 |
| | December | 11139 | 10366.042 | 302.261111 | 470.697222 |

| Year | Month | | | | |
|---|---|---|---|---|---|
| **2013** | January | 11379 | 10492.458 | 63.611111 | 822.930556 |
| | February | 10809 | 10589.042 | -5.280556 | 225.238889 |
| | March | 10499 | 10668.750 | -22.672222 | -147.077778 |
| | April | 10164 | 10774.125 | 55.419444 | -665.544444 |
| | May | 11091 | 10897.458 | -65.030556 | 258.572222 |
| | June | 10731 | 11005.417 | -325.772222 | 51.355556 |
| | July | 10672 | 11076.750 | -293.705556 | -111.044444 |
| | August | 10255 | 11150.917 | -419.822222 | -476.094444 |
| | September | 10893 | 11295.083 | 34.677778 | -436.761111 |
| | October | 11776 | 11526.625 | 247.344444 | 2.030556 |
| | November | 12103 | 11787.458 | 428.969444 | -113.427778 |
| | December | 12247 | 12094.708 | 302.261111 | -149.969444 |
| **2014** | January | 11983 | 12486.500 | 63.611111 | -567.111111 |
| | February | 11985 | 12952.292 | -5.280556 | -962.011111 |
| | March | 12783 | 13496.792 | -22.672222 | -691.119444 |
| | April | 13437 | 14031.333 | 55.419444 | -649.752778 |
| | May | 14078 | 14553.542 | -65.030556 | -410.511111 |
| | June | 15118 | 15099.958 | -325.772222 | 343.813889 |
| | July | 15688 | 15659.500 | -293.705556 | 322.205556 |
| | August | 16418 | 16263.833 | -419.822222 | 573.988889 |
| | September | 17798 | 16855.250 | 34.677778 | 908.072222 |
| | October | 17700 | 17364.333 | 247.344444 | 88.322222 |
| | November | 18712 | 17794.542 | 428.969444 | 488.488889 |
| | December | 18752 | 18126.208 | 302.261111 | 323.530556 |
| **2015** | January | 18907 | 18391.083 | 63.611111 | 452.305556 |
| | February | 19565 | 18625.167 | -5.280556 | 945.113889 |
| | March | 19397 | 18710.583 | -22.672222 | 709.088889 |
| | April | 19041 | 18693.417 | 55.419444 | 292.163889 |
| | May | 18799 | 18687.625 | -65.030556 | 176.405556 |
| | June | 18357 | 18662.125 | -325.772222 | 20.647222 |
| | July | 18806 | | -293.705556 | |
| | August | 18918 | | -419.822222 | |
| | September | 17348 | | 34.677778 | |
| | October | 17738 | | 247.344444 | |
| | November | 18535 | | 428.969444 | |
| | December | 18317 | | 302.261111 | |

**Observations:**

1. From Table 1, we observe that the Seasonal components for the Auto sector indices are positive during the period September-January, with the highest value occurring in the month of November. The Seasonal component becomes the minimum in the month of June every year. The Trend values have consistently increased over the period 2010 – 2015. However, the rate of growth of the Trend value has decreased during January 2015- June 2015 and possibly even after that period. The Random component has shown considerable fluctuations in its values. However, the Trend being the predominant component in the overall time series, the time series is quite amenable for forecasting.

2. It is natural for the auto sector to have a dominant seasonal component as purchase of vehicles coincide, both with the religious festivals, and also during the third quarter as during this period the economic activity starts to pick up. The results from the agricultural sector adds to the seasonality as the impact of the rains are available from September onwards.

## 5. RESULTS OF FORECASTING

In this Section, we discuss some forecasting methods that we have applied on the time series data of the data in the Auto sector index. We have followed four different approaches in forecasting which have produced different levels of accuracy in forecasting. However, the main observation from our analysis is that the Auto sector index is very much dominated by the Trend and the Seasonal components of the time series, with the Random component generally playing a minor role.

The different methods for forecasting, and a method for understanding the roles played by the constituent components of the time series that we have followed, are as follows.

**Method 1:** The time series data of the Auto sector index from January 2010 to December 2014 are used for forecasting the monthly indices for the year 2015. The forecasting is made at the end of December 2014. Error is forecasting is also computed for each month in order to have an idea about the accuracy in forecasting technique.

**Method II:** Forecasting for the monthly indices for the year 2015 is made on the basis of time series data from January 2010 till the end of the previous month for which the forecast is made. For example, for the purpose of forecasting the monthly index for March 2015, time series data from January 2010 till February 2015 are considered. As in Method 1, error in forecasting is also computed.

**Method III:** In this method, we first use the time series data for the Auto sector monthly indices from January 2010 to December 2014 to compute its Trend and Seasonal components. This method yields Trend components from July 2010 to June 2014 with the Trend values for the first six months and last six months being truncated. Based on the Trend values till Jun 2014, we make forecasts for the Trend values during January 2015 to June 2015. We add the forecasted Trend values to the Seasonal components of the corresponding months (based on the time series data from January 2010 to December 2014) to arrive at the forecasted aggregate of the Trend and Seasonal components. Now we consider the full time series of the Auto sector indices from January 2010 to December 2015 and decompose it into its Trend, Seasonal and random components. We compute the aggregate of the actual Trend and the actual Seasonal component values. Finally, to have an idea about forecasting accuracy, we compute the percentage of deviation of the actual aggregate of Trend and Seasonal component values with their corresponding forecasted aggregate values for each month during January 2015 to June 2015.

**Method IV:** In this method, we first consider the time series of the Auto sector month indices during January 2010 to December 2014. The time series is decomposed into its Trend, Seasonal and Random components and we compute the aggregate of the Trend and the Seasonal components during July 2010 to June 2014. Note that the Trend components from January 2010 to June 2010 and also from July 2014 to December 2014 will not be available after the decomposition. Next, we consider the time series data from January 2011 to December 2015. We again compute the aggregate of the Trend and the Seasonal components for the new time series (i.e., the time series from January 2011 to December 2015). As in Method 3, in order to have an idea about the accuracy in the forecasting process, we find the percentage of deviation of the computed aggregate of Trend and Seasonal components for each month during June 2011 to July 2014, computed based on the two time series (January 2010 – December 2014 and January 2011 – December 2015).

**RESULTS**

**Method I:** As mentioned earlier in this Section, we forecast for each month in January 2015 based on time series data from January 2010 to December 2014. We use **HoltWinters** function in R library "forecast" for this purpose. In order to make a robust forecasting, we use HoltWinters model with a Trend and an additive Seasonal component that best fits the Auto index time series data. The forecast "horizon" in the HoltWinters model has been chosen to be 12 so that the forecasted values for all months of 2015 can be obtained.

**Table 1: Computation Results using Method I**

| Month (A) | Actual Value (B) | Forecasted Value (C) | Error Percentage (C-B)/B *100 |
|---|---|---|---|
| January | 18907 | 18507.47 | -2.11 |
| February | 19565 | 17988.12 | -8.05 |
| March | 19397 | 18383.55 | -5.22 |
| April | 19041 | 19365.45 | 1.70 |
| May | 18799 | 19545.82 | 3.97 |
| June | 18357 | 19544.08 | 6.47 |
| July | 18806 | 19872.57 | 5.67 |
| August | 18918 | 20349.92 | 7.57 |
| September | 17348 | 21221.72 | 22.33 |
| October | 17738 | 21926.94 | 23.62 |
| November | 18535 | 22486.62 | 21.32 |
| December | 18317 | 22685.94 | 23.85 |

**Observations:** We observe from Table 2 that the forecasted values closely match the actual values even when the forecast horizon is long (12 months). This clearly shows that HoltWinters model with Trend and additive Seasonal components is very effective in forecasting Auto sector monthly indices during the period 2010 -2015. The error values during September – December 2015 are comparatively larger due to sudden downward movement in the index. This has surely affected the trend component of the time series which was impossible to predict in December 2014.

**Method II:** As mentioned earlier in this Section, in this approach we forecast the Auto sector index for each month in 2015 by taking into account time series data till the month before the month of forecast. We use HoltWinters model with additive Seasonal component using a prediction horizon of 1. Since the prediction horizon is small, the model can capture any possible change in Trend and Seasonal components much more effectively. The only factors that can induce error in forecasting are: (i) appreciable change in the Seasonal component, (ii) a very strong and abruptly changing random component. Following this approach, we attempt to get a deeper insight into the behavior of the Auto sector index time series in terms of the behavior of its Seasonal and Random components.

**Table 2: Computation Results using Method II**

| Month (A) | Actual Value (B) | Forecasted Value (C) | Error Percentage (C-B)/B *100 |
|---|---|---|---|
| January | 18907 | 18507 | -2.12 |
| February | 19565 | 18426 | -5.82 |
| March | 19397 | 19825 | 2.21 |
| April | 19041 | 20307 | 6.65 |
| May | 18799 | 19687 | 4.72 |
| June | 18357 | 18937 | 3.16 |
| July | 18806 | 18609 | -1.05 |
| August | 18918 | 19077 | 0.84 |
| September | 17348 | 19794 | 14.10 |
| October | 17738 | 18038 | 1.70 |
| November | 18535 | 18104 | -2.33 |
| December | 18317 | 18470 | 0.84 |

**Observations:** We observe from Table 3 that the forecasted values very closely match the actual values in this approach. This clearly demonstrates that HolWinters additive model with a prediction horizon of 1 is absolutely capable of capturing the behavior of the Auto sector time series indices. The high error value in the month of September can be very well attributed to an abnormal behavior of the Seasonality component during that month.

**Method III:** In the earlier part of this Section, we have already discussed the approach followed in this method. We have used the time series data of the Auto sector indices from January 2010 to December 2015 to compute the actual values of the Trend and Seasonal. However, since the actual values of Trend component are not available for the period July 2015 – December 2015, we concentrate only on the period January 2015 to June 2015 for the purpose of forecasting . The actual Trend and Seasonal component values and their aggregated monthly values are noted in Columns B, C and D respectively in Table 3. Now using the time series data during January 2010 to December 2014, the Trend and Seasonal components are recomputed. Since the Trend values during July 2014 to December 2014 will not be available after this computation, we make a forecast for the trend values for the period January 2015 to June 2015 using HoltWinters method of forecast. The forecasted Trend values and the past Seasonal component values and their corresponding aggregate values are noted in columns E, F and G respectively in Table 3. The error values are also computed.

The major objective of this approach is to investigate how the Trend and the Seasonal components in the times series data exhibit their behavior and how strong they are in terms of influencing the overall times series values. Since the values of the Random component cannot be forecasted effectively, a time series which is dominated by its Trend and Seasonality components will be more amenable to forecasting. In this approach, we investigate how efficiently we can forecast the aggregate values of the Trend and Seasonality components of Auto sector index time series.

Table 3: Computation Results using Method III

| Month | Actual Trend | Actual Seasonal Component | Actual (Trend + Seasonal) | Forecasted Trend Component | Past Seasonal Component | Forecasted (Trend + Seasonal) | % Error |
|---|---|---|---|---|---|---|---|
| A | B | C | D | E | F | G | (G-D)/D *100 |
| Jan | 18391 | 64 | 18455 | 15700 | 61 | 15761 | -14.60 |
| Feb | 18625 | -5 | 18620 | 15736 | -131 | 15605 | -16.19 |
| Mar | 18710 | -23 | 18687 | 15771 | -90 | 15681 | -16.09 |
| Apr | 18693 | 55 | 18748 | 15803 | 93 | 15896 | -15.21 |
| May | 18688 | -65 | 18623 | 15835 | 1 | 15836 | -14.97 |
| Jun | 18662 | -326 | 18336 | 15987 | 221 | 16056 | -12.43 |

**Observation:** We observe that error values are quite moderate considering the fact that we had to make a forecast of the Trend values over a long horizon (12 months). We forecasted the trend values for the period January 2015 to June 2015 on the basis of the Trend component values till June 2014. Any possible change in the actual Trend values or variations in Seasonality component will induce error in this approach. Considering these aspect, we note that the resultant error margins are quite moderate which emphasizes the fact that the time series is mainly dominated by the Trend and Seasonality and hence it is amenable to HoltWinters forecasting model with Trend and additive Seasonality components.

**Method IV:** The objective of this method is to gain an insight into the role played by the Trend and the Seasonal components of the time series on the overall auto sector aggregate index. As we mentioned earlier in this Section, this approach is based on comparison of the aggregate of the

Trend and Seasonal components of a time series over two different period of time. First, we construct a time series using the data during January 2010 to December 2014, and then compute the Trend and Seasonal components and their aggregate values. We refer to this computation as "Computation 1". The Trend, Seasonal and their aggregate values in Computation 1 are noted in columns A, B and C respectively in Table 4. Next, construct the second time series using the data during January 2011 to December 2015 and repeat the computation of the Trend, Seasonal and their aggregate values. The Trend, Seasonal and their aggregate values in Computation 2 are noted in columns D, E and F respectively in Table 4. The percentages of variation of the aggregate values in both computations are noted for each month during July 2011 to June 2014.

The main essence of this approach is to investigate how the Trend and Seasonal components change over period of time and how their aggregate values change when a portion of time series is substituted by another. We made Computation 1 on the basis of time series data from January 2010 to December 2014, while Computation 2 was made using time series data from January 2011 to December 2015. If there is a structural difference between the time series data in 2010 and 2015, then we expect that difference to be reflected in the aggregate Trend and Seasonal values. If any significant difference is observed in the aggregate values of Trend and Seasonal components for the period 2011 – 2014, then we can conclude that the time series is dominated by its random components and hence it is not amenable for forecasting.

**Table 4: Computation Results using Method IV**

| Year | Month | Computation 1 | | | Computation 2 | | | % Variation |
|---|---|---|---|---|---|---|---|---|
| | | **Trend** | **Seasonal** | **Sum** | **Trend** | **Seasonal** | **Sum** | |
| | | A | B | C = A + B | D | E | F = D + E | (F - C)/C *100 |
| **2011** | Jul | 8792 | -264 | **8498** | 8792 | -258 | **8534** | 0.36 |
| | Aug | 8812 | -452 | **8360** | 8812 | -477 | **8335** | -0.30 |
| | Sep | 8916 | -82 | **8834** | 8916 | -19 | **8897** | 0.71 |
| | Oct | 9000 | 336 | **9336** | 9000 | 134 | **9134** | -2.16 |
| | Nov | 9057 | 417 | **9474** | 9057 | 325 | **9382** | -0.97 |
| | Dec | 9100 | 332 | **9432** | 9100 | 195 | **9295** | -1.45 |
| **2012** | Jan | 9135 | 61 | **9196** | 9135 | 77 | **9212** | 0.17 |
| | Feb | 9190 | -131 | **9059** | 9190 | 214 | **9404** | 3.80 |
| | Mar | 9281 | -90 | **9191** | 9281 | 118 | **9399** | 2.26 |
| | Apr | 9390 | 93 | **9483** | 9390 | 21 | **9411** | -0.76 |
| | May | 9527 | 1 | **9528** | 9527 | -40 | **9487** | -0.43 |
| | Jun | 9720 | -221 | **9499** | 9720 | -289 | **9431** | -0.72 |
| | Jul | 9952 | -264 | **9688** | 9952 | -258 | **9694** | -0.06 |
| | Aug | 10108 | -452 | **9656** | 10108 | -477 | **9631** | -0.26 |
| | Sep | 10168 | -82 | **10086** | 10168 | -19 | **10149** | 0.62 |
| | Oct | 10181 | 336 | **10517** | 10181 | 134 | **10315** | -1.92 |
| | Nov | 10236 | 417 | **10653** | 10236 | 325 | **10561** | -0.86 |
| | Dec | 10366 | 332 | **10698** | 10366 | 195 | **10561** | -1.28 |
| **2013** | Jan | 10492 | 61 | **10553** | 10492 | 77 | **10569** | 0.15 |

|      | Feb | 10582 | -131 | **10451** | 10589 | 214 | **10803** | 3.37 |
|      | Mar | 10669 | -90 | **10579** | 10669 | 118 | **10787** | 1.97 |
|      | Apr | 10774 | 93 | **10867** | 10774 | 21 | **10795** | -0.66 |
|      | May | 10897 | 1 | **10898** | 10897 | -40 | **10857** | -0.37 |
|      | Jun | 11005 | -221 | **10784** | 11005 | -289 | **10716** | -0.63 |
|      | Jul | 11077 | -264 | **10813** | 11077 | -258 | **10819** | 0.06 |
|      | Aug | 11151 | -452 | **10699** | 11151 | -477 | **10674** | -0.23 |
|      | Sep | 11295 | -82 | **11213** | 11295 | -19 | **11276** | 0.56 |
|      | Oct | 11527 | 336 | **11863** | 11526 | 134 | **11660** | -1.71 |
|      | Nov | 11787 | 417 | **12204** | 11787 | 325 | **12112** | -0.75 |
|      | Dec | 12095 | 332 | **12427** | 12095 | 195 | **12290** | -1.10 |
| **2014** | Jan | 12487 | 61 | **12548** | 12487 | 77 | **12564** | 0.13 |
|      | Feb | 12952 | -131 | **12821** | 12952 | 214 | **13166** | 2.69 |
|      | Mar | 13497 | -90 | **13407** | 13497 | 118 | **13615** | 1.55 |
|      | Apr | 14031 | 93 | **14124** | 14031 | 21 | **14052** | -0.51 |
|      | May | 14553 | 1 | **14554** | 14553 | -40 | **14513** | -0.28 |
|      | Jun | 15100 | -221 | **14879** | 15100 | -289 | **14811** | -0.46 |

**Observation:** It is clear that sum of the Trend and Seasonal component are consistently same for the period July 2011 to Jun 2014. It may be noted that we could not make computations during January 2011 to June 2011 and also during July 2014 to December 2014 due to non-availability of Trend values during those periods. So the change of the time series due to substitution of the 2010 data by 2015 data has virtually no impact on the Trend and Seasonal components. The effect of Random component being very minor, the time series is quite amenable to forecasting.

## 6. RELATED WORK

Forecasting of daily stock prices has attracted considerable attention from the research community. Neural network based approaches have been proposed to make various kind of forecasting. Mostafa used neural network technique to predict stock market movements in Kuwait [3]. Kimoto et al applied neural network-based approach using historical accounting data and various macroeconomic parameters to forecast variations in stock returns [4]. Leigh et al used liner regression and simple neural network models for prediction stock market indices in the New York Stock Exchange for the period 1981-1999 [5]. Hammad et al have demonstrated that artificial neural network (ANN) model can be trained so that it converges while maintaining high level of precision in forecasting of stock prices [6]. Dutta et al used ANN models for forecasting Bombay Stock Exchange's SENSEX weekly closing values for the period of January 2002-December 2003 [7]. Ying et al used Bayesian network (BN)-approach to forecast stock prices of 28 companies listed in DJIA during 198-1998. Tsai and Wang showed results that highlighted the fact that Bayesian Network-based approaches have better forecasting ability than traditional regression and Neural Network-based approaches [8]. Tseng et al applied traditional time series decomposition (TSD), HoltWinters (H/W) models, Box-Jenkins (B/J) methodology and Neural

Network- based approach to 50 randomly selected stocks from September 1, 1998 to December 31, 2010 with an objective of forecasting future stock prices [9]. They have observed that forecasting errors are lower for B/J, H/W and normalized Neural network model, while the errors are appreciably large for time series decomposition and non-normalized Neural Network model. Moshiri and Cameron [10] designed a Back Propagation Network (BPN) with econometric models to forecast inflation using (i) Box-Jenkins Autoregressive Integrated Moving Average (ARIMA) model, (ii) Vector Autoregressive (VAR) model and (ii) Bayesian Vector Autoregressive (BVAR) model. Datta Chaudhuri and Ghosh presented Artificial Neural Network (ANN) models based on various back propagation algorithms for the purpose of predicting volatility in the Indian stock market through volatility of NIFTY returns and volatility of Gold returns [11].

In contrast to the work mentioned above, our approach in this paper is based on structural decomposition of a time series to study the behavior of the auto sector in India during 2010-2015 and we have proposed various forecasting techniques. We have computed the relative accuracies of each of the forecasting techniques and also critically analyzed under what situations a particular technique performs best. The analysis can be used as a broad approach for forecasting the behavior of other stock market indices in India. However, our results also elicit one point clearly – the auto sector index in India had a relatively large random component during 2015 that made forecasting task quite difficult during this year

## 7. CONCLUSION

In this work, we have analyzed the auto index time series in India during the period January 2010 to December 2015. We have used R programming language to structurally decompose the time series values in three components- Trend, Seasonal, and Random. The decomposition results have provided us a deeper insight into the behavior of the Auto index time series. Based on the results we have been able to identify the months during which the Seasonal component plays a major role. We have also been able to have an idea about the Trend of the Auto sector index. Using these decomposition results, we have proposed four approaches for forecasting the index values of the Auto sector with a forecast horizon as large as 12 months. We have also introduced a technique to understand the structural analysis of the time series data using its Trend and Seasonal components. The forecast results clearly demonstrate the effectiveness and efficiency of our proposed forecasting techniques. Even in presence of a very dominant Random component in time series values, our techniques have been able to achieve quite an acceptable level of forecasting accuracies.

The results obtained from the above analysis is extremely useful for portfolio construction. When we perform this analysis for other sectors as well, it will help portfolio managers and individual investors to identify which sector, and in turn which stock, to buy/sell in which period. It will also help in identifying which sector, and hence which stock, is dominated by the random component and thus is speculative in nature.